\documentclass[a4paper,aps,pra,showpacs,showkeys,preprintnumbers]{revtex4-2}
\usepackage[utf8]{inputenc}
\usepackage[english]{babel}
\usepackage[T2A]{fontenc}
\usepackage{amssymb,amsfonts,amsmath,mathtext,enumerate,float,dsfont}
\usepackage{dblfloatfix}
\usepackage{graphics,epsfig,epstopdf}
\usepackage{caption}
\usepackage{cmap}
\usepackage{multirow}
\usepackage{indentfirst}
\usepackage[usenames]{color}
\usepackage{amsthm}
\usepackage{physics}
\usepackage{xcolor}
\usepackage{type1ec}% 
\usepackage{dcolumn}%
\usepackage{dblfloatfix}
\usepackage{adjustbox}
\usepackage[graphicx]{realboxes}
\usepackage{hyperref}

\DeclareMathOperator{\arccosh}{ArcCosh}

\begin{document}

%%%%%%%%%%%%%%%%%
\def\BY{\begin{eqnarray}}
\def\EY{\end{eqnarray}}
\def\L{\label}
\def\nn{\nonumber}
\def\ds{\displaystyle}
\def\o{\overline}
\def\({\left (}
\def\){\right )}
\def\[{\left [}
\def\]{\right]}
\def\<{\langle}
\def\>{\rangle}
\def\h{\hat}
\def\td{\tilde}
\def\r{\vec{r}}
\def\ro{\vec{\rho}}
\def\h{\hat}
\def\v{\vec}
%%%%%%%%%%%

\title{Bosonic quantum error correction using squeezed Fock states}

\author{E. N. Bashmakova}
\author{S. B. Korolev}
\author{T. Yu. Golubeva } 
\affiliation{St.Petersburg State University, Universitetskaya nab. 7/9, St.Petersburg, 199034, Russia}

\begin{abstract}
In the paper, we develop a bosonic quantum error correction code based on squeezed Fock states. We compare our proposed code with one based on squeezed Schrödinger's cat states using the Knill-Laflamme cost function and the Petz map fidelity. We demonstrate that squeezed Fock states are competitive in protecting information in a channel with particle loss and dephasing.
\end{abstract}
\maketitle

\section{Introduction}
One of the main goals of quantum informatics and quantum optics is to build universal quantum computing. Currently, quantum computing is in the so-called NISQ (Noisy Intermediate Scale Quantum) era \cite{Preskill2018QuantumCI}. This era is marked by the development of computing systems with a limited number of logic units and no error correction procedure. When developing intermediate-scale computing, the primary concern is reducing errors during computations associated with imperfections of the physical systems and the impossibility of completely isolating them from the environment. However, error suppression is not enough to build a full-scale computing procedure. Error correction is necessary to prevent errors from accumulating as operations are performed. Quantum error correction (QEC) codes detect and correct errors. The principle of QEC codes is to encode information using specific quantum states with certain properties. These properties are determined by the types of errors being corrected. This approach to QEC refers to bosonic quantum codes (BQC) \cite{Schlegel_2022,Grimsmo2020,Terhal_2020,Wallraff2004,PhysRevLett.111.120501,PhysRevA.94.042332,PhysRevA.64.012310,PhysRevA.93.012315,Mirrahimi_2014,PhysRevA.59.2631,PhysRevA.75.042316,PhysRevA.70.022317,PhysRevA.97.032346} where information is redundantly encoded in quantum oscillator states.

Error correction protocols have been proposed for various quantum states at present. For example, encoding information by GKP (Gottesman-Kitaev-Preskill) states \cite{Vasconcelos:10, PhysRevA.64.012310}  allows one to construct a protocol that protects against small phase-space displacement errors. GKP states have a periodic sharp-peaked grid structure in both of the quadratures. If there are phase-space displacement errors, the state structure changes. The magnitude of the displacement is a critical factor in the error correction protocol. In this case, the error correction is possible if the magnitude of the displacement is smaller than the threshold value. Otherwise, the error cannot be corrected. At the same time, the particle loss error can be corrected using QEC codes based on GKP states \cite{PhysRevA.108.052413}. Despite some success of such an error correction protocol \cite{PhysRevA.73.012325, Eaton_2019}, the generation of such states in the optical range remains a challenging task.

Schrödinger's cat states \cite{Sychev2017,Buzek1995} are another example of quantum states applied to QEC. These states are a superposition of two coherent states, ($|\alpha\rangle$ and $|-\alpha\rangle$). QEC codes based on these states can protect information from particle loss and dephasing errors. For effective protection, cat states with amplitude $\alpha = 2$ or higher are required \cite{Ralph_2003,Hastrup_2022}. Currently, several protocols have been proposed to generate states similar to Schrödinger's cat states \cite{Sychev2017,Ourjoumtsev_cat2007, Polzik2006, Huang2015, Ulanov2016, Gerrits2010, Takahashi2008, Baeva2022,Podoshvedov_2023,Thekkadath2020}. However, the achievable values in the optical range ($|\alpha| \leq 1.9$) \cite{Sychev2017, Ourjoumtsev_cat2007, Huang2015, Ulanov2016, Gerrits2010, Takahashi2008} are insufficient for error correction protocols to work. In the microwave range, Schrödinger's cat states with high fidelity and large $\alpha$ amplitude values can be generated \cite{He2023,PhysRevA.99.022302}. However, in such systems, there are other difficulties related to the always-on Kerr nonlinearity suppression, which prevents individual control of logic states \cite{KirchmairG2013Ooqs}.

From the point of view of QEC codes, the squeezed Schrödinger's cat (SSC) states \cite{Grimsmo2020} are of interest. Based on SSC states, an error correction code was developed that is capable of simultaneously correcting two types of errors: dephasing errors and particle loss errors. Unlike traditional Schrödinger's cat states, SSC states should have a large squeezing degree and a small amplitude $\alpha$ for further incorporation into QEC protocols. Protocols for generating SSC state are proposed and discussed in \cite{Wang2022,Bashmakova_2023,PhysRevLett.114.193602,PhysRevA.106.043721,Ourjoumtsev2007}. However, the question of efficient generation of these quantum states with high fidelity and probability remains open.

Squeezed Fock (SF) states \cite{Olivares_2005} are an example of well-studied non-Gaussian states. To the authors' knowledge, there are no yet error correction protocols based on these states. SF states have an advantage over other non-Gaussian states in terms of their experimental realization. The generation of the different types of Schrödinger's cat states, in most cases, is carried out only approximately \cite{PhysRevLett.132.230602,Podoshvedov_2023, Takase2021}. We have proposed a protocol for generating exact (with fidelity equal to 1) SF states \cite{PhysRevA.109.052428}.

This paper is devoted to the demonstration of a QEC code for particle loss and dephasing errors that is based on SF states. We compare our protocol to one that uses SSC states. We discuss two measures: the modified KL (Knill–Laflamme) cost function constructed in terms of Kraus operators \cite{Korolev_qec} and the Petz map fidelity \cite{PRXQuantum.3.020314,PhysRevA.86.012335,PhysRevA.81.062342}. The recovery operation has been explicitly specified and can always be experimentally implemented in the proposed QEC protocol, which is based on SF states \cite{PhysRevResearch.6.043034}.

\section{Measure for comparing codewords in bosonic quantum codes}
\subsection{Approximate error correction code}\label{general_A}

In this section, we briefly review the fundamentals of bosonic quantum codes (BQC)\cite{Schlegel_2022,Grimsmo2020,Terhal_2020,Wallraff2004,PhysRevLett.111.120501,PhysRevA.94.042332,PhysRevA.64.012310,PhysRevA.93.012315,Mirrahimi_2014,PhysRevA.59.2631,PhysRevA.75.042316,PhysRevA.70.022317,PhysRevA.97.032346}.

Consider a physical system with Hilbert space denoted by $\mathcal{H}$. This system will be used to transmit and process quantum information. QEC codes primarily involve an encoding procedure, which can be described by a linear invertible transformation $\mathcal{C}: \mathcal{B}(\mathcal{H}) \rightarrow \mathcal{B}(\mathcal{H}_{d})$ defined on a $d$-dimensional Hilbert subspace $\mathcal{H}_{d}$ of the physical space $\mathcal{H}$. The effect of errors on the quantum system is described by a completely positive trace-preserving (CPTP) mapping $\mathcal{N}: \mathcal{B}(\mathcal{H}_{d})\rightarrow \mathcal{B}(\mathcal{H}_{n})$. The $\mathcal{N}$ map can describe the action of different types of errors in the quantum channel, such as particle loss or dephasing errors. The complete positivity of the $\mathcal{N}$ map means that the map is decomposed in terms of Kraus
operators $\{\hat{K}_{j}\}$ as \cite{KRAUS1971311}
\begin{align}
\mathcal{N}(\hat{\rho})=\sum_{j=0}^{\infty} \hat{K}_{j} \hat{\rho}\hat{K}_{j}^{\dagger}.
\label{N}
\end{align}

In terms of Kraus operators, the trace-preserving map $\mathcal{N}$ is defined by the condition
\begin{align}
\sum_{j=0}^{\infty} \hat{K}_{j}^{\dagger} \hat{K}_{j}=\mathcal{I},
\label{trace_save}
\end{align}
where $\mathcal{I}$ is the identity operation defined on the same Hilbert space as the map $\mathcal{N}$. To correct the errors, we perform a recovery procedure. It is described by the CPTP map $\mathcal{R}: \mathcal{B}(\mathcal{H}_{n})\rightarrow \mathcal{B}(\mathcal{H}_{d})$.

Next, we perform the decoding procedure $\mathcal{C}^{-1}: \mathcal{B}(\mathcal{H}_{d}) \rightarrow \mathcal{B}(\mathcal{H})$.

For few-photon systems in discrete variables, the set of Kraus operators is limited by the dimension of space. We will consider continuous variable quantum systems. In this case, there is no simplification. Physical grounds are needed to limit the number of Kraus operators under consideration.

Only particle loss and dephasing errors in the quantum channel are caused by the channel's interaction with the vacuum reservoir. In terms of Kraus operators, the particle loss errors are described as \cite{Leviant2022quantumcapacity,PhysRevA.97.032346}
\begin{align}
\hat{K}^{(1)}_{j}=\sqrt{\frac{\gamma_{1}^{j}}{j!}}(1-\gamma_{1})^{\hat{n}/2}\hat{a}^{j},
\label{photon_loss}
\end{align}
where $\gamma_{1}=1-e^{-\kappa_{1}t}\in[0,1]$ is the dimensionless rate of particle loss (determined by the rate $\kappa_{1}$ over a time interval $t$). Here, $\hat{a}$ stands for the annihilation operator of the quantum field that obeys the canonical commutation relation $[\hat{a}, \hat{a}^{\dagger}]=1$. $\hat{n}\equiv\hat{a}^{\dagger}\hat{a}$ is the excitation number operator. Each Kraus operator corresponds to the $j$-excitation loss event. Notice that for $j=0$ and $\gamma_{1}\neq 0$ the action of the noise channel $\mathcal{N}$ is not identical. This is due to the factor $\exp(-\frac{\gamma_{1}}{2}\hat{n})$. It is known as the back action or no-jump evolution \cite{PhysRevA.97.032346}.

Realistic bosonic modes are also subject to bosonic dephasing errors \cite{Leviant2022quantumcapacity}. It can be defined in terms of Kraus operators as
\begin{align}
\hat{K}^{(2)}_{j}=\sqrt{\frac{\gamma_{2}^{j}}{j!}}\exp(-\frac{\gamma_{2}}{2}\hat{n}^{2})\hat{n}^{j}.
\label{def}
\end{align}
Here, the dimensionless dephasing rate $\gamma_{2}=\kappa_{2}t$ is defined in terms of the dephasing rate $\kappa_{2}$ over a time interval $t$. The state is smeared in the phase space as a result of the dephasing error. The Kraus operators of the dephasing error are functions of $\hat{n}$ only. They are diagonal in the Fock basis. Consequently, the dephasing channel does not introduce transitions in comparison to the particle loss channel.

Quantum systems can be subjected to both particle loss and dephasing errors in “real” noisy channels. The quantum channels $\mathcal{N}(\hat{\rho})[\gamma_{1}]$ and $\mathcal{N}(\hat{\rho})[\gamma_{2}]$ commute \cite{Leviant2022quantumcapacity}. It allows us to study the impact of these errors on quantum states independently of each other:
\BY
\mathcal{N}(\hat{\rho})[\gamma_{1},\gamma_{2}]=\mathcal{N}(\hat{\rho})[\gamma_{1}]\circ\mathcal{N}(\hat{\rho})[\gamma_{2}]=\mathcal{N}(\hat{\rho})[\gamma_{2}]\circ\mathcal{N}(\hat{\rho})[\gamma_{1}],
\EY

Using the explicit form of the Kraus operators for particle loss and dephasing errors (\ref{photon_loss})-(\ref{def}), we can introduce a set of elementary errors $\hat{E}$ of the following form
\begin{align} \label{Error_set}
    \mathcal{E}=\left\{\hat{I}, \hat{a},\hat{a}^{2}, ...,   \hat{a}^{\dagger} \hat{a}, \left(\hat{a}^{\dagger} \hat{a}\right)^{2}, ..., \left(\hat{a}^{\dagger} \hat{a}\right)^{j}\right\}.
\end{align}
Thus, to protect the information in the channel from these types of errors is associated with the ability to correct the following set of elementary errors $\mathcal{E}$, or a subset $\left\{\hat{E}\right\} \in \mathcal{E}$.

In the BQC codes, we consider the code space $\mathcal{H}_{d}$ to be a two-dimensional subspace of an infinite-dimensional Hilbert space $\mathcal{H}$. The code space $\mathcal{H}_{d}$ is characterized by two basis states (codewords) that encode the logical states $|0_{L}\rangle$ and $|1_{L}\rangle$. When individual errors act on codewords, they move to other states from the error subspace. The quantum information is not lost but rather encoded in the error subspace when an error occurs in a quantum channel. It allows us to detect, thus correct, errors without degrading the quantum information in the channel. All these requirements can be written in the form of KL conditions \cite{PhysRevA.55.900}:
\begin{align}
\left\langle i_{L}\left|\hat{E}_{a}^{\dagger} \hat{E}_{b}\right| j_{L}\right\rangle =\delta_{i j} \alpha_{a b},
\label{KL}
\end{align}
where $ i,j\in \lbrace 0,1\rbrace$,  and $\hat{E}_{a}, \hat{E}_{b}\in \mathcal{E}$, and $ \alpha_{a b}$ is a matrix that does not depend on $i$ and $j$. $\delta_{i j}$ is the Kronecker symbol. The presented condition is a necessary and sufficient condition for the recovery of codewords after the action of the errors. In this case, the KL conditions determine whether codewords can protect information from a certain set of errors $\left\{\hat{E}\right\}$. If the KL conditions are exactly satisfied, the QEC code is the perfect code. For example, the code for correcting single excitation losses based on binomial states is perfect \cite{PhysRevA.97.032346}. However, such states are a mathematical abstraction and are not experimentally realizable. BQC that are experimentally realizable are all approximate codes. For them, the KL conditions are satisfied approximately. These codes will be further discussed. 

In the following section, we will discuss measures that allow us to study the degree of sensitivity of the codewords to different types of errors.

\subsection{KL cost function }
As a measure for comparing codewords, we will use the KL cost function based on the conditions (\ref{KL}). This function is proposed in \cite{Schlegel_2022,Reinhold} and used to compare two codes in the paper \cite{Schlegel_2022,Korolev_qec}. The KL cost function is given by the following expression:
\begin{align}
    C_{KL} \left(\left\lbrace \hat{E}\right\rbrace\right)=\sum _{a,b}\left| f_{00ab}-f_{11ab}
\right|^2+\left| f_{01ab}\right|^2,
\label{KL_def}
\end{align}
where the KL tensor is given by the expression
\begin{align}
    f_{ijab}=\left\langle i_L \left|\hat{E}^\dag _a\hat{E}_b\right|j_L\right\rangle,
\end{align}
and $\lbrace\hat{E}\rbrace$ is represented by a set of error operators over which the sum is taken. In the limit of perfect code, for which the KL conditions (\ref{KL}) are exactly satisfied, the value of the KL cost function is equal to zero. The larger the KL cost function value, the worse the approximate code corrects errors.

Note that the KL cost function is defined by a set of elementary errors $\left\{\hat{E}\right\}$. In this case, in the definition (\ref{KL}) all elementary errors are included with equal weights. In reality, each error has its accumulation rate $\gamma_{i} \;(i=\{1,2\})$. To estimate the KL cost function, given the accumulation rate of individual errors, we need to use the Kraus operators $\hat{K}$ instead of the elementary error operators $\hat{E}$. Let us define the KL cost function for a noisy channel in terms of Kraus operators as:
\begin{align}
    \widetilde{C}_{KL} \left(\left\lbrace \hat{K}\right\rbrace\right)=\sum _{i,j}\left| \left\langle 0_L \left|\hat{K}_i^\dagger \hat{K}_j\right|0_L\right\rangle-\left\langle 1_L \left|\hat{K}_i^\dagger \hat{K}_j\right|1_L\right\rangle
\right|^2+\left|\left\langle 0_L \left|\hat{K}^\dagger _i\hat{K}_j\right|1_L\right\rangle\right|^2.
\label{KL_K}
\end{align}

As a result,  the KL cost function defined in this way has the same properties as the original one. Specifically, this function tends to zero for the perfect code. It is important to note that the analysis based on the KL cost function provides only a qualitative assessment of the codewords. The KL cost function estimates only the encoding ($\mathcal{C}$) and noise channels ($\mathcal{N}$) of the QEC procedure described in Section \ref{general_A}. The QEC code is also dependent on the recovery map ($\mathcal{R}$). Thus, it is required to construct both the encoding and recovery procedures. It is necessary to calculate the channel fidelity for quantifying the error correction performance of the given code.

\subsection{Channel fidelity and optimal recovery operation}\label{general_B}
Let us consider the general quantum channel $\mathcal{Q}: \mathcal{B}(\mathcal{H})\rightarrow \mathcal{B}(\mathcal{H})$, $\mathcal{Q}= \mathcal{C}^{-1}\circ\mathcal{R}\circ\mathcal{N}\circ\mathcal{C}$, described in Section \ref{general_A}, consists of the encoding ($\mathcal{C}$), noise ($\mathcal{N}$), recovery ($\mathcal{R}$) and decoding ($\mathcal{C}^{-1}$) maps. We assume the encoding map ($\mathcal{C}$) is the perfect. The decoding map ($\mathcal{C}^{-1}$) does not introduce errors if codewords are chosen to be orthogonal. This is the case we will consider below.

The goal of the QEC protocol is to construct an optimal recovery map $\mathcal{R}$ that minimizes the difference between the input state and state restored after the noise map $\mathcal{N}$. This problem can be mathematically described as the determination of the minimum distance between $\mathcal{Q}$ and $\mathcal{R}\circ \mathcal{N}$ channels. The channel fidelity $F$ allows quantifying the error correction performance of the given code  \cite{Schlegel_2022,PhysRevA.97.032346,PhysRevA.54.2614}. For the general quantum channel $\mathcal{Q}$, the channel fidelity $F$ is defined as
\begin{align}
F=\frac{1}{4}\sum_{j}{|\Tr\{\hat{Q}_{j}\}|^2},
\label{Fopt}
\end{align}
where $\hat{Q}_{j}$ is the $i$-th the Kraus operators of the quantum channel $\mathcal{Q}(\hat{\rho})=\sum_{j} \hat{Q}_{j} \hat{\rho}\hat{Q}_{j}^{\dagger}$. Substituting the operators $\hat{Q}_{j}$ into the Eq.(\ref{Fopt}) directly in terms of the Kraus operators of the noise $\mathcal{N}$ and recovery $\mathcal{R}$ maps, we obtain that the channel fidelity has the following form 
\begin{align}
F=\frac{1}{4}\sum_{r,j}{|\Tr\{\hat{R}_{r}\hat{K}_{j}\}|^2}.
\label{Fopt2}
\end{align}
Here, $\hat{R}_{r}$ is the Kraus operator of the recovery map $\mathcal{R} (\hat{\rho})=\sum_{r} \hat{R}_{r} \hat{\rho}\hat{R}_{r}^{\dagger}$.

\begin{figure}[H]
   \centering
   \includegraphics[scale=0.55]{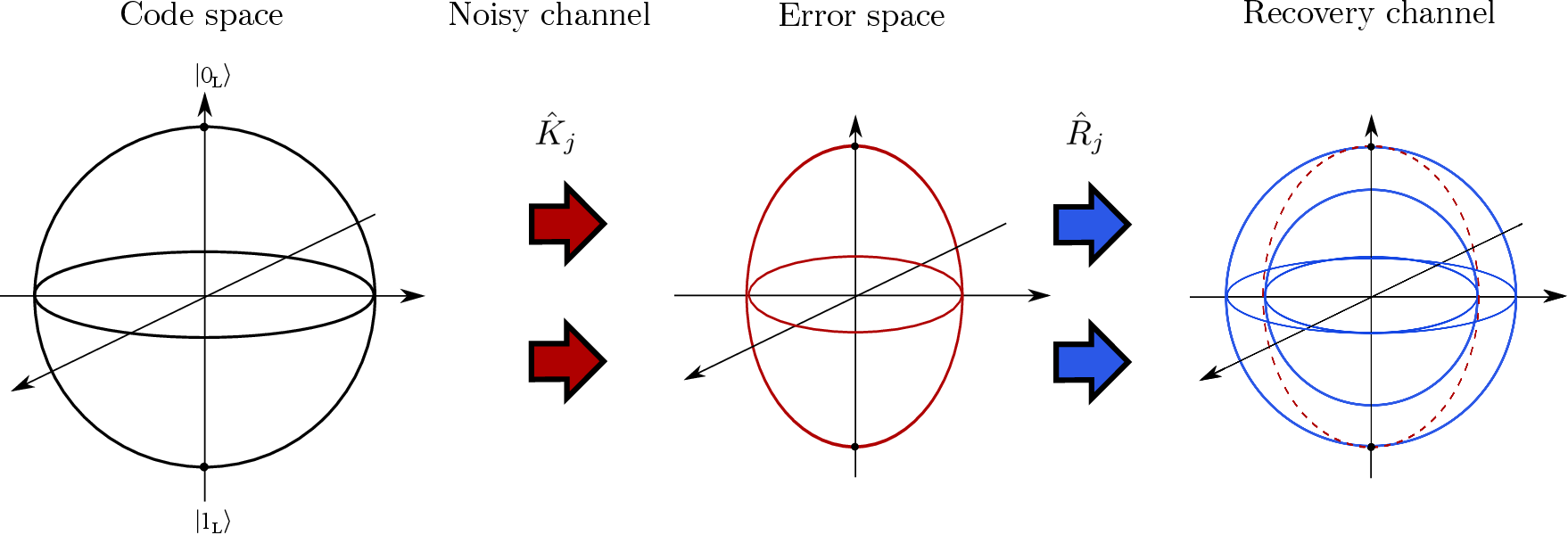}
    \caption{Schematic illustration of the effect of errors $\hat{K}_{j}$ on codewords $|0_{L}\rangle$ and $|1_{L}\rangle$. The encoding of quantum information in the two-dimensional subspace $\mathcal{H}_{d}$ of the Hilbert space $\mathcal{H}$ is shown as a black solid Bloch sphere. Codewords $|0_{L}\rangle$ and $|1_{L}\rangle$ lie respectively at the north and south poles of the Bloch sphere of the logical qubit. The action of the noise map (marked as red arrows) is shown as a red ellipsoid. In this case, codewords are no longer poles of the ellipsoid. The action of the recovery map (marked as blue arrows) is demonstrated as two blue spheres, the radii of which correspond to the long and short axes of the red ellipsoid. Between the two spherical surfaces, there is a figure that corresponds to the optimal recovery procedure.}
   \label{fig:Bloch}
\end{figure}

The “optimal”{} channel fidelity $F_{opt}$ is associated with the optimization of the recovery procedure. To find the optimal recovery map $\mathcal{R}_{opt}$, we use the quantum process tomography methods. So we expand the Kraus recovery operators in the basis of projection operators $\{\hat{B}^{(m)}_{j}\}$ \cite{Schlegel_2022,Kosut2009} as
\begin{align}
\hat{R}_{r}=\sum_{j}\sum_{m=1}^{4}{x_{r,j}\hat{B}^{(m)}_{j}},
\label{R}
\end{align}
where the set of operators $\{\hat{B}^{(m)}_{j}\in\mathcal{H}_{d}\otimes\mathcal{H}_{n}\subset \mathcal{B} (\mathcal{H}_{n})\}$ describes how the quantum state outside the code space is projected back onto the code space. Here, $x_{r,j}$ are some complex coefficients that describe the corresponding Kraus recovery operator. The superscript $(m)$ is responsible for the type of the operator $\hat{B}^{(m)}_{j}$. Then the search for the optimal recovery procedure consists of finding, for a given set of projection operators $\{\hat{B}^{(m)}_{j}\}$, coefficients $x_{r,j}$ that maximize $F_{opt}=F(\mathcal{R}_{opt}\circ\mathcal{N})$ the channel fidelity.

Let us discuss in more detail the choice of projection operators $\{\hat{B}^{(m)}_{j}\}$.  The state of the logical qubit $a|0_{L}\rangle+b|1_{L}\rangle$ can be represented as a vector on the Bloch sphere (Fig.\ref{fig:Bloch}). In this case, as shown in Fig.\ref{fig:Bloch}, codewords $|0_{L}\rangle$ and $|1_{L}\rangle$ lie, respectively, at the north and south poles of the Bloch sphere under consideration. Particle loss and dephasing errors transform codewords into states in the error subspace. In Fig.\ref{fig:Bloch}, the subspace of such states is depicted as an ellipsoid. The effect of noise can be shown by small changes in the lengths of the basis vectors. In general, the figure representing the error subspace is not an ellipsoid. It can be any isomorphic map of the logical space. The Kraus recovery operators $\hat{R}_{r}$ in (\ref{R}) are defined via the basis of mutually orthogonal recovery operations. Any arbitrary recovery operator $\hat{R}_{r}:\mathcal{H}_{n}\rightarrow\mathcal{H}_{d}$ can be uniquely decomposed into operators $\{\hat{B}^{(m)}_{j},\;m\in[1,4],\; j=0,..,N\}$, where $N$ is the number of orthogonal error subspaces under consideration. Any arbitrary rotation on the Bloch sphere is described using the Pauli operators in the code space: $\{\hat{\sigma}_{x}, \hat{\sigma}_{y}, \hat{\sigma}_{z}, \hat{I}\}$. Thus, we represent the set of projection operators $\{\hat{B}^{(m)}_{j}\}$ in the following form
\begin{align}
&\hat{B}_{j}^{(1)}=\hat{I}\hat{K}_{j}^{\dagger}=\ket{0_{L}}\bra{\psi_{j}^{+}}+\ket{1_{L}}\bra{\psi_{j}^{-}},
\\&\hat{B}_{j}^{(2)}=\hat{\sigma}_{x}\hat{K}_{j}^{\dagger}=\ket{0_{L}}\bra{\psi_{j}^{-}}+\ket{1_{L}}\bra{\psi_{j}^{+}},
\\&\hat{B}_{j}^{(3)}=\hat{\sigma}_{y}\hat{K}_{j}^{\dagger}=i\ket{0_{L}}\bra{\psi_{j}^{-}}-i\ket{1_{L}}\bra{\psi_{j}^{+}},
\\&\hat{B}_{j}^{(4)}=\hat{\sigma}_{z}\hat{K}_{j}^{\dagger}=\ket{0_{L}}\bra{\psi_{j}^{+}}-\ket{1_{L}}\bra{\psi_{j}^{-}}.
\label{B}
\end{align}
Here, $\ket{\psi_{j}^{+}}=\hat{K}_{j}|0_{L}\rangle$ and $\ket{\psi_{j}^{-}}=\hat{K}_{j}|1_{L}\rangle$ states 
are orthogonal codewords affected by particle loss or dephasing errors of the noise map $\mathcal{N}$. To obtain the final error subspaces spanned by the orthogonal states, it is necessary to orthogonalize the states using Gram-Schmidt orthogonalization \cite{Schlegel_2022}. Note that the choice of the orthogonal basis of codewords is not unique. Within each error subspace, it is determined up to a rotation.

The Eq. (\ref{Fopt}) is the definition of channel fidelity in terms of Kraus operators. But for direct calculations, it is easier to use Choi matrices to define channel fidelity \cite{Schlegel_2022,Kosut2009}:
\begin{align}
F=\frac{1}{4}\sum_{i,j}{[X]_{i,j}[W]_{i,j}}.
\label{Fopt2}
\end{align}
Here, $[X]_{i,j}$ is the Choi matrix of the recovery map. $[W]_{i,j}$ is the process matrix, which is defined as
\begin{align}
&[X]_{i,j}=\sum_{i,j}{x_{r,i}x^{*}_{r,j}},\label{Fopt_m1}
\\& [W]_{i,j}=\sum_{l}\sum_{m,n}{\Tr\{\hat{B}^{(m)}_{i}\hat{K}_{l}\}\Tr\{\hat{B}^{(n)}_{j}\hat{K}_{l}\}^{*}},
\label{Fopt_m2}
\end{align}
where  $\Tr$ is the trace of the superoperators. The CPTP properties of the general quantum channel $\mathcal{Q}$ correspond to the matrices $X$ and $W$ being positive semidefinite. A detailed discussion of the relationship between Eq. (\ref{Fopt}) and Eq. (\ref{Fopt2}) is presented in Appendix \ref{Fidelity_channel}.

Note that the optimal recovery map $\mathcal{R}_{opt}$ must also have the properties of a CPTP map. Therefore, the trace-preserving Kraus recovery operators $\hat{R}_{r}$ must be satisfied
\BY
\sum_{r}\hat{R}_{r}^{\dagger}\hat{R}_{r}=\sum_{m,n}\sum_{i,j}[X]_{i,j}{(\hat{B}^{(m)})^{\dagger}}_{i}\hat{B}^{(n)}_{j}=\mathcal{I}.
\EY

It should be noted that Eq. (\ref{Fopt2}) is linear in $[X]_{i,j}$. The matrix $X$ is positive semidefinite. Then the calculation of the channel fidelity can be written as an optimization problem of the convex semidefinite program (SDP) \cite{PhysRevLett.94.080501,PhysRevA.75.012338}:
\begin{align}\label{SDP1}
\text{maximize} \quad&\frac{1}{4}\sum_{i,j}{[X]_{i,j}[W]_{i,j}}=\frac{1}{4}\Tr\{X W\}\nonumber
\\ \text{subject to}\quad &\sum_{i,j}\sum_{m,n}[X]_{i,j}{(\hat{B}^{(m)})^{\dagger}}_{i}\hat{B}^{(n)}_{j}=\hat{I}
\\& X\succeq 0,\nonumber
\end{align}
where the parameters $[X]_{i,j}$ form a convex set. Solving the SDP is an optimal solution of the Choi matrix $X$ that maximizes the channel fidelity and gives an explicit form of the recovery map $\mathcal{R}_{\text{opt}}$.

The optimal Choi matrix $X^{\text{opt}}$ is the solution to the optimization problem (\ref{SDP1}). We can obtain an explicit form of the optimal recovery map $\hat{R}^{\text{opt}}$ by the following decomposition \cite{Schlegel_2022}
\begin{align}
& X^{opt}=V D V^{\dagger},
\\&\hat{R}^{opt}_{r}=\sqrt{d}_{r}\sum_{i}\sum_{m}V_{ir}\hat{B}^{(m)}_{i}.
\end{align}
Here, $V$ is a unitary matrix with elements $V_{ir}$. $D$ is a diagonal matrix with elements $d_{r}$.

Therefore, calculating the channel fidelity is the SDP optimization problem from the NP (nondeterministic polynomial time) complexity class. This means that calculational complexity is found to grow exponentially with an increase in problem dimension. Currently, there are many solvers available to solve the above SDP
\cite{Kosut2009,10.1007/978-1-84800-155-8_7}.
Next, we will discuss what measures allow us to avoid solving the optimization problem.

\subsection{Petz map fidelity}\label{FPetz_D}
In the previous subsection, we have discussed in detail the KL conditions. It is necessary and sufficient conditions for the recovery of codewords after the action of the errors. In other words, the KL conditions only assume that there is a recovery channel. To obtain an explicit form of the optimal recovery operation, it is necessary to calculate the channel fidelity. As shown above, it is associated with solving the complex optimization problem. In this section, we discuss another measure that allows quantifying the error correction performance of a given code but does not assume the solution of the optimization problem.

In \cite{PRXQuantum.3.020314} authors considered a fixed recovery map $\mathcal{R}_{\text{Petz}}$. The recovery map is constructed using only one projector $\hat{P}_{L}=|0_{L}\rangle\langle0_{L}|+|1_{L}\rangle\langle 1_{L}|$, in contrast to $\mathcal{R}_{opt}$.
The recovery procedure $\mathcal{R}_{\text{Petz}}$ is described by a set of Kraus operators $\mathcal{R}_{\text{Petz}}(\hat{\rho})=\sum_{r} \hat{R}^{\text{Petz}}_{r} \hat{\rho}\;(\hat{R}^{\text{Petz}}_{r})^{\dagger}$ of the following form
\begin{align}
\hat{R}^{\text{Petz}}_{r}=\hat{P}_{L}\hat{K}^{\dagger}_{r}\mathcal{N}\big(\hat{P}_{L}\big)^{-1/2},
\end{align}
Here, $\mathcal{N}\big(\hat{P}_{L}\big)=\sum_{m,\mu}\hat{K}_{m}\ket{\mu_{L}}\bra{\mu_{L}}\hat{K}^{\dagger}_{m}$, where $\ket{\mu_{L}}$ is a codeword.  It can be argued that it is applicable when errors in a quantum channel lead to small changes in the lengths of the basis vectors. Such errors weakly deform the Bloch sphere. Otherwise, the recovery operation will be rendered ineffective.

The recovery map $\mathcal{R}_{\text{Petz}}$ determines the channel fidelity $F_{\text{Petz}}$. It has the following form:
\begin{align}
F_{Petz}=\frac{1}{4}||\Tr_{L}\sqrt{(m^{-1}\otimes \mathcal{I}_{r})M}||_{F}^{2},
\label{FPetz}
\end{align}
where the KL tensor is given by the expression
\begin{align}
M_{[\mu,l],[\nu,n]}=\bra{\mu_{L}}\hat{K}^{\dagger}_{l}\hat{K}_{n}\ket{\nu_{L}}.
\end{align} 
Here, $\mathcal{I}_{r}$ is the $r\times r$ identity matrix, $r$ is the number of Kraus operators of the noise map $\mathcal{N}$, and $||A||_{F}=\sqrt{\Tr(A^{*}A)}$ is the Frobenius norm. The partial trace operation in Eq. (\ref{FPetz}) is  the operation $\big(\Tr_{L}A\big)_{l,k}=\sum_{\mu}A_{[\mu,l],[\mu,k]}$. Considering the potential non-orthogonality of codewords, the elements of the matrix $m$ are defined as $m_{\mu,\nu}=\bra{\mu_{L}}\ket{\nu_{L}}$.

It is important to note that $F_{\text{Petz}}$ is related with the optimal channel fidelity $F_{\text{opt}}$. Thus, $F_{\text{Petz}}$ provides a two-sided bound on $F_{\text{opt}}$ as \cite{PRXQuantum.3.020314}:
\begin{align}
F_{\text{Petz}} \leq F_{\text{opt}} \leq \frac{1}{2}(1+F_{\text{Petz}}).
\end{align}
This interval for $F_{\text{opt}}$ is also shown in Fig.\ref{fig:Bloch} as the space between two spherical blue surfaces. The optimal value of $F_{\text{opt}}$ is situated within the interval between two spherical blue surfaces. The optimal value of $F_{\text{opt}}$ is located within a narrower interval as the value of $F_{\text{Petz}}$ approaches one. Thus, by obtaining a sufficiently large value of $F_{\text{Petz}}$, we can accurately estimate the optimal channel fidelity, avoiding computational challenges.

\section{Particle loss and dephasing errors correction protocol based on squeezed Fock states}\label{SF_protocol}

In the previous section, we have discussed in detail the measures characterizing the performance of QEC protocols. We have demonstrated that there are two ways to compare the proposed protocol with one based on SSC states. Firstly, the KL cost function (\ref{KL_K}) is based on Kraus error operators. Secondly, the Petz map fidelity $F_{\text{Petz}}$ is shown by Eq. (\ref{FPetz}).  The KL cost function assesses the quality of codewords without considering the recovery operation. This function assumes that the perfect recovery operation can be carried out. It does not introduce errors into the recovery result. As a result, the recovery procedure for noisy channels with small errors is considered by the channel fidelity $F_{\text{Petz}}$. Our goal is to develop a particle loss and dephasing error correction protocol based on SF states. For this protocol, we present the results of the KL cost function and the channel fidelity $F_{\text{Petz}}$ calculations. Next, we compare the obtained result with  QEC protocols with SSC states. This comparison is motivated by the fact that the QEC code using SSC states can correct particle loss and dephasing errors simultaneously \cite{Grimsmo2020}. In addition, this code is similar to our proposed code in terms of experimental realization. Both SF states and states close to SSC states can be obtained experimentally (in optical schemes with measurements) \cite{Bashmakova_2023,PhysRevA.109.052428}.

\begin{figure}[H]
   \centering
   \includegraphics[scale=0.25]{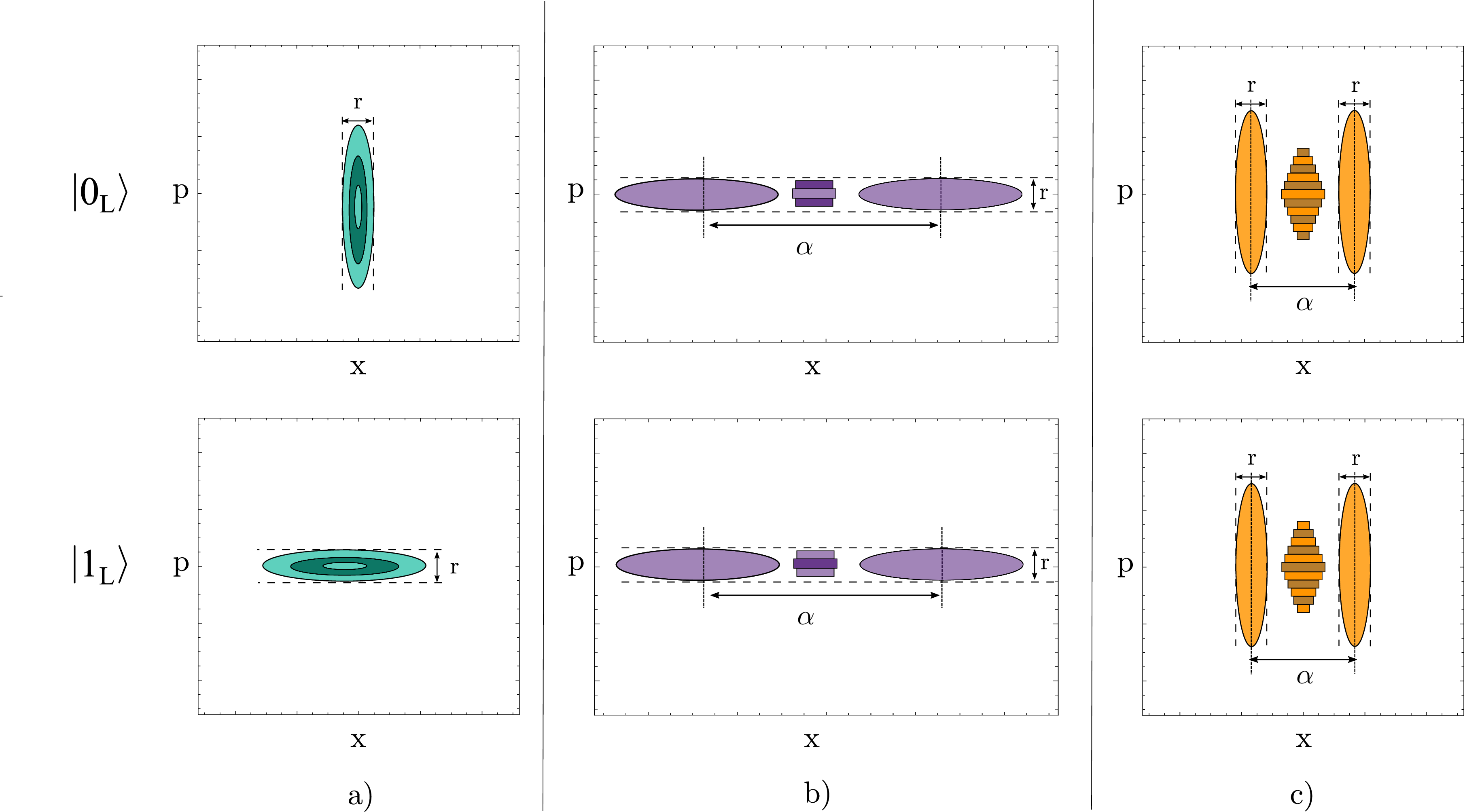}
    \caption{Schematic representation of phase portraits of codewords $|0_{L}\rangle$ and $|1_{L}\rangle$, respectively: a) the second squeezed Fock state with squeezing parameters $r$ and $-r$; b) even and odd squeezed Schrödinger’s cat states with squeezing parameters $r$, when coherent states $\ket{\alpha}$ and $\ket{-\alpha}$ are spaced along the axis and squeezed along the orthogonal one; c) even and odd squeezed Schrödinger’s cat states with squeezing parameters $r$, when coherent states $\ket{\alpha}$ and $\ket{-\alpha}$ are spaced and squeezed along the axis.}
   \label{fig:KL}
\end{figure}

The QEC protocol for particle loss and dephasing errors will be built using SF states, which are defined as follows:
\begin{align}
    |r,n\rangle=\hat{S}(r)|n\rangle,
\end{align}
where $\hat{S}(r)$ is the squeezing
operator, $r$ is the squeezing parameter, and $|n\rangle$ is the n-photon Fock state. To correct for particle loss and dephasing errors, we require states with specific parity \cite{PhysRevLett.111.120501,PhysRevA.94.042332} and structure in phase space \cite{Schlegel_2022,Leviant2022quantumcapacity}. Fig. \ref{fig:KL}a illustrates the phase portrait schematically, demonstrating that the SF states satisfy all these properties. 
The SF states have a definite parity. It guarantees the sensitivity of such states to small photon losses. The phase portrait of such states has an asymmetry on the phase plane. The preferred direction ensures that such states are sensitive to dephasing errors.

To correctly compare QEC protocols, we also need to fix the mean number of photons ($\langle n \rangle$) in the quantum states that are used as codewords. By fixing these parameters, it will be possible to compare states with similar energy properties. As will be shown below, the mean number of photons $\langle n \rangle$ in quantum states affects the number of Kraus error operators (\ref{N}).

Let us discuss which SF states should be considered as codewords $|0_{L}\rangle$ and $|1_{L}\rangle$. We are interested in orthogonal SF states. In this case, the decoding procedure ($\mathcal{C}^{-1}$) does not introduce errors into the quantum channel $\mathcal{Q}$, discussed in sections \ref{general_A} and \ref{general_B}. However, SF states are only orthogonal for certain values of the parameters $r$ and $n$. In addition, we aim to select the minimum number $n$ of states with the desired properties. This is because with an increase in the number $n$, the probability of generating such states decreases \cite{PhysRevA.109.052428}. Consequently, SF states with the minimum number of $n=2$ are the ones for which a squeezing parameter $r$ can be chosen to guarantee that the SF states are orthogonal. It is achieved only for a pair of SF states with squeezing parameter $r=\pm0.57$. The mean number of photons in such states is $\langle n \rangle=3.83$. A detailed discussion of the choice of the SF states that will be used as codewords is presented in Appendix \ref{SF_codewords}. Thus, to construct the protocol for correcting particle loss and dephasing errors, we will use the following SF states
\begin{align}
&|0_{L};2\rangle=\hat{S}\left(0.57\right)|2\rangle,\label{SF0}\\
&|1_{L};2\rangle=\hat{S}\left(-0.57\right)|2\rangle.\label{SF1}
\end{align}

Let us compare our code with the code based on SSC states, which are defined as
\begin{align}
    &|0_{L,SSC}\rangle=\frac{1}{N_{+}}\left(|\alpha,r\rangle+|-\alpha,r\rangle\right),\\
    &|1_{L,SSC}\rangle=\frac{1}{N_{-}}\left(|\alpha,r\rangle-|-\alpha,r\rangle\right),
\end{align}
where $|\alpha,r\rangle=\hat{S}\left(r\right)\hat{D}\left(\alpha\right)|0\rangle$ are squeezed coherent states, and $N_{\pm}=\sqrt{2\left(1\pm e^{-2|\alpha| ^2 e^{2r}}\right)}$ is a normalization factor. These states are characterized by the amplitude $\alpha$ and squeezing parameter $r$. With a mean number of photons of $\langle n \rangle=3.83$, there is an infinite number of SSC states. To investigate the relationship between the amplitude and squeezing parameter of the SSC state, we will fix two amplitude values $\alpha=0.5$ and $\alpha=1.0$ \cite{Grimsmo2020,Hastrup_2022}. With these amplitude values $\alpha$, we will choose the squeezing parameter values $r$ so $\langle n \rangle=3.83$. Note that the squeezing parameter $r$ and the corresponding amplitudes $\alpha$ are required for two classes of SSC states (see Table \ref{tab:state}). The first class of the SSC states, the phase portraits of which are presented in Fig. \ref{fig:KL}b, include states such that the coherent states $\ket{\alpha}$ and $\ket{-\alpha}$ are "spaced" along one quadrature, and squeezed along the orthogonal one (correspond to the symbol $\alpha^{\perp}$). For another class of states (Fig. \ref{fig:KL}c), it is characteristic that the coherent states $\ket{\alpha}$ and $\ket{-\alpha}$ are "spaced" and squeezed along one quadrature (correspond to the symbol $\alpha^{\|}$) \cite{Wang2022}. 
At the same time, both classes of SSC states are used in quantum error correction protocols \cite{Lau2019,Xu2023,Grimsmo2020}. Note that for a given value of the amplitudes $\alpha$ and squeezing parameter $r$, SSC states of different parities are orthogonal. Thus, we will compare the second SF state with the set of four SSC states in terms of their ability to correct particle loss and dephasing errors.

\begin{table}[h]
\begin{center}
\caption{\label{tab:state} Parameters of the considered quantum non-Gaussian states with the mean number of photons $\langle n \rangle=3.83$, which will be used to compare the photon loss and dephasing error correction protocols: squeezed Schrödinger cat states (SSC) and the second squeezed Fock state (SF). The following symbols are used: $ \alpha^{\perp}$ is the amplitude of the SSC, for which it is typical that the coherent states $\ket{\alpha}$ and $\ket{-\alpha}$ are "spaced" along one quadrature, and squeezed along the orthogonal one (Fig.  \ref{fig:KL}b); $ \alpha^{\|}$ is the amplitude of the SSC, for which it is typical that the coherent states $\ket{\alpha}$ and $\ket{-\alpha}$ are "spaced" and squeezed along one quadrature (Fig. \ref{fig:KL}c).}
\begin{tabular}{|c|c|c|}
\hline
Symbol of SCS states  &SSC states   & SF state                  \\ \hline
$\alpha^{\|}_{0.5}$& $\alpha=0.5$, $r=-1.41$     & \multirow{4}{*}{n=2, r=0.57} \\ \cline{1-2}
$\alpha^{\|}_{1.0}$&$\alpha=1.0$, $r=-1.36$ &                         \\ \cline{1-2}
$\alpha^{\perp}_{0.5}$& $\alpha=0.5$, $r=1.39$     &                         \\ \cline{1-2}
$\alpha^{\perp}_{1.0}$& $\alpha=1.0$, $r=1.29$     &                         \\ \hline
\end{tabular}
\end{center}
\end{table}

We have discussed the quantum states that we use as codewords. Now, let us analyze the physical parameters that describe the quantum channel. We will start with a discussion of the noise quantum channel $\mathcal{N}$ considered in Section \ref{general_A}. As we noted in the previous section, each noise channel $\mathcal{N}$ is characterized by the set of Kraus error operators $\{\hat{K}_{j}\}$. The set of Kraus error operators is infinite for quantum systems in continuous variables. This is determined by the upper limit of the sum in Eq. (\ref{N}). To limit the number of Kraus operators under consideration, there must be physical grounds. Such reasons can be related to the two parameters. Firstly, it is the mean number of photons $\langle n \rangle$ in quantum states that are used as codewords. Secondly, it is associated with the low value of the dimensionless rate $\gamma$ that describes the noise channel. The trace-preserving property of Kraus operators (\ref{trace_save}), which we discussed in section \ref{general_A}, is significant in this context. The property is approximately satisfied by the truncation of the series of particle loss $\hat{K}_{j}^{(1)}$ and dephasing errors $\hat{K}_{j}^{(2)}$ by Kraus operators: 
\begin{align}
&\sum_{j} \langle \mu_{L} |\big(\hat{K}_{j}^{(1)}\big)^{\dagger} \hat{K}^{(1)}_{j}|\mu_{L} \rangle{}\approx 1+\mathcal{O}(\gamma^{2}_{1}\langle n^{2} \rangle)\geq1,\label{rad1}\\
&\sum_{j}\langle \mu_{L} |\big(\hat{K}^{(2)}_{j}\big)^{\dagger} \hat{K}^{(2)}_{j}|\mu_{L} \rangle{}\approx 1+\mathcal{O}(\gamma^{2}_{2}\langle n^{4} \rangle)\geq1.\label{rad2}
\end{align}
It is important to note that the accuracy of the trace-preserving property of the Kraus operators is influenced by the mean number of photons $\langle n \rangle$ in codewords $|\mu_{L} \rangle{}$ and the low value of the dimensionless error rate $\gamma$.  We define a set of operators $\hat{K}^{(1)}_{j}$ and $\hat{K}^{(2)}_{j}$ that we will use to develop QEC protocols by fixing the second order of expansion $(\gamma_{1}^{2}\langle n^{2} \rangle)$ and $(\gamma_{2}^{2}\langle n^{4} \rangle)$ of the series in Kraus operators of the particle loss and  dephasing errors, respectively:
\begin{align}
&\hat{K}^{(1)}_{1}=\hat{I}-\frac{\gamma_{1}}{2} \hat{a}^{\dagger} \hat{a}, \label{OK_pl1}\\
&\hat{K}^{(1)}_{2}=\sqrt{\gamma_{1}} \hat{a}\label{OK_pl2},\\
&\hat{K}^{(2)}_{1}=\hat{I}-\frac{\gamma_{2}}{2}\left(\hat{a}^{\dagger} \hat{a}\right)^{2}, \label{OK_deph1}\\
&\hat{K}^{(2)}_{2}=\sqrt{\gamma_{2}} \hat{a}^{\dagger} \hat{a} \label{OK_deph2}.
\end{align}
Note that the various degrees of $\langle n \rangle$ indicate that such cutting off will be correct for different values of the error rates $\gamma$ for the considered types of errors.

\begin{figure}[H]
   \centering
   \includegraphics[scale=0.5]{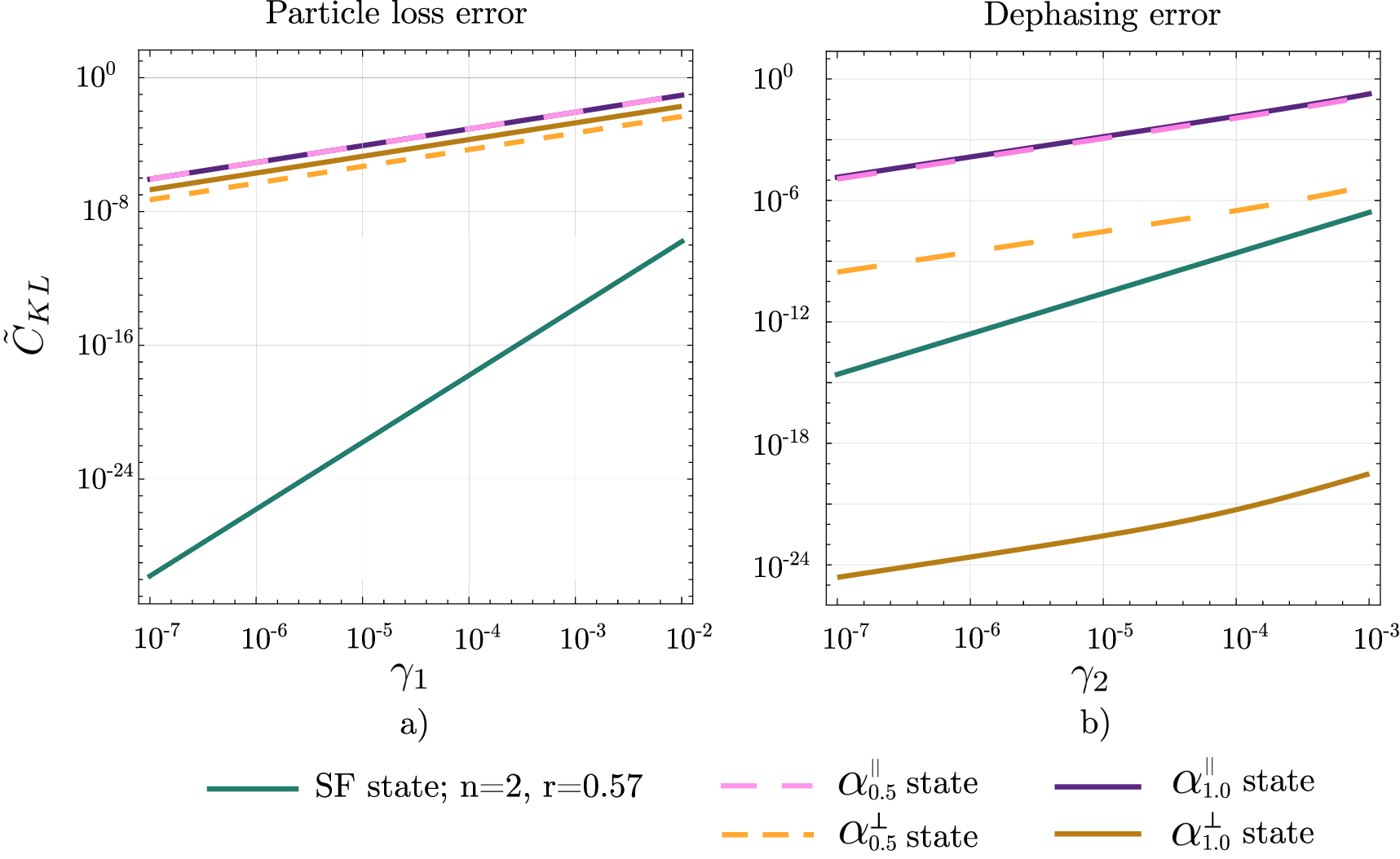}
    \caption{Dependence of the KL cost function on: a) the rate of particle loss error $\gamma_{1}$ for the set of errors $\lbrace \hat{I}, \hat{a}, \hat{a}^\dagger \hat{a}\rbrace$; b) the rate of dephasing error $\gamma_{2}$ for the set of errors $\lbrace \hat{I},\left(\hat{a}^\dagger \hat{a}\right)^2\rbrace$. The following quantum states were investigated (see Table \ref{tab:state}): 1) squeezed Schrödinger's cat states: $\alpha_{0.5}^{\|}$ (marked in pink color), $\alpha_{0.5}^{\perp}$ (marked in orange color), $\alpha_{1.0}^{\|}$ (marked in purple color), $\alpha_{1.0}^{\perp}$ (marked in brown color); 2) the second squeezed Fock state with the squeezing $r=0.57$ (marked in green color). The mean number of photons in the considered states is $\langle n \rangle=3.83$.}
   \label{fig:KL_plot}
\end{figure}

\begin{figure}[H]
   \centering
   \includegraphics[scale=0.5]{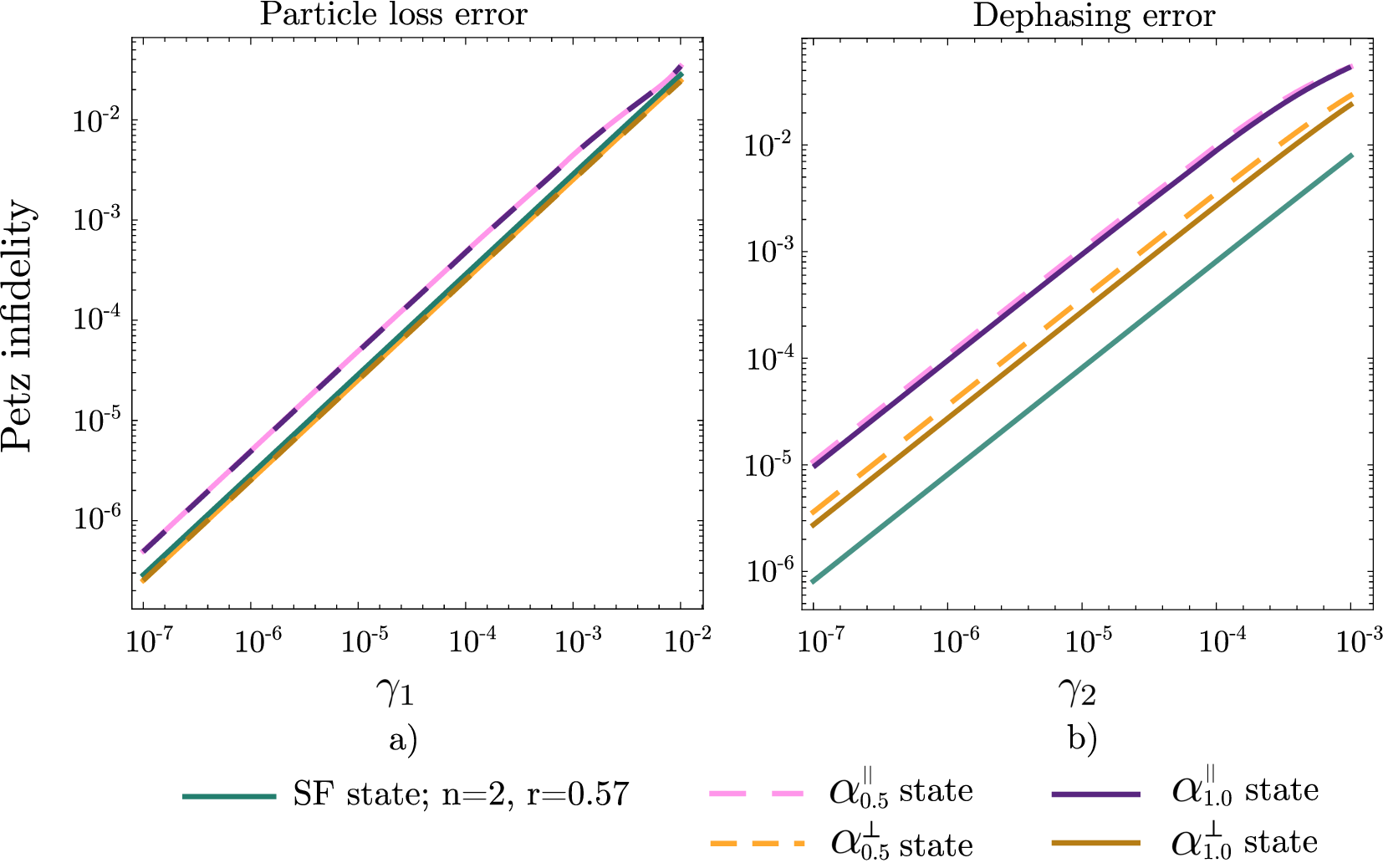}
    \caption{Dependence of the Petz map infidelity of a channel on: a) the rate of particle loss error $\gamma_{1}$ for the set of errors $\lbrace \hat{I}, \hat{a}, \hat{a}^\dagger \hat{a}\rbrace$; b) the rate of dephasing error $\gamma_{2}$ for the set of errors $\lbrace \hat{I},\left(\hat{a}^\dagger \hat{a}\right)^2\rbrace$. The following quantum states were investigated (see Table \ref{tab:state}): 1) squeezed Schrödinger's cat states: $\alpha_{0.5}^{\|}$ (marked in pink color), $\alpha_{0.5}^{\perp}$ (marked in orange color), $\alpha_{1.0}^{\|}$ (marked in purple color), $\alpha_{1.0}^{\perp}$ (marked in brown color); 2) the second squeezed Fock state with the squeezing $r=0.57$ (marked in green color). The mean number of photons in the considered states is $\langle n \rangle=3.83$.}
   \label{fig:fidelity_FPetz}
\end{figure}

Let us discuss the QEC code for particle loss and dephasing errors using the second SF state. Our results on the KL cost function (Fig. \ref{fig:KL_plot}) and Petz map fidelity (Fig. \ref{fig:fidelity_FPetz}) demonstrate that the proposed approximate QEC protocol based on the second SF state can correct errors in a channel with the low value of the dimensionless photon loss rates ($\gamma_{1}\in[10^{-7};10^{-2}]$) and dephasing rates ($\gamma_{2}\in[10^{-7};10^{-3}]$). For the noise channel with high particle loss error rates ($\gamma_{1}>10^{-2}$) and dephasing error ones ($\gamma_{2}>10^{-3}$), the set of the Kraus error operators (\ref{rad1})-(\ref{rad2}) is not applicable. In other words, to describe a noise channel, it is not enough to use only the Kraus error operators from the set (\ref{OK_pl1})-(\ref{OK_deph2}). It is necessary to use the following expansion orders (\ref{N}). However, the correction results do not show significant change when studying low-noise channels with the error rates $\gamma_{1,2}<10^{-7}$. As a result, we looked at this error rate limit value. It should be noted that the error range we study is physically significant. For instance, in \cite{Hann2025}, the authors discuss error rate ranges of $\gamma_1 \in [10^{-6}; 10^{-2}]$, $\gamma_2 \in [10^{-8}; 10^{-4}]$ for the implementation of surface codes constructed using Schrödinger cat states. The authors of \cite{Putterman2025} report error rates on the order of $10^{-3}$. In \cite{PhysRevB.94.014506}, photon loss rates in the channel were achieved at approximately $10^{-3}$.

Based on SF and SSC states, we will compare the QEC protocols for particle loss and dephasing errors. This comparison will be done with the KL cost function and the Petz map fidelity.

Firstly, we use the KL cost function (Fig. \ref{fig:KL_plot}) to compare codewords that are being considered.  Fig. \ref{fig:KL_plot}a shows that the second SF state, marked in green color, corrects particle loss errors significantly better than the considered SSC states. The KL cost function for the second SF state is smaller than for any SSC state over the entire considered interval of the dimensionless rate $\gamma_{1}$. As for the dephasing error (Fig. \ref{fig:KL_plot}b), the KL cost function curve for the SF state (marked in green color) lies between the corresponding ones for SSC states with different amplitudes $\alpha$. For large values of $\alpha$, the code using the SSC state corrects this type of error better than the second SF state. But as $\alpha$ increases, the SSC state code performs a correction to particle loss error worse (Fig. \ref{fig:KL_plot}a). %Generally, the second SF state is more effective in correcting particle loss and dephasing errors due to the combination of these two factors.

The SSC's code against photon loss and dephasing errors with the squeezing parameter $r<0$ will be addressed separately. The KL cost function values for the states $\alpha_{0.5}^{\|}$ and $\alpha_{1.0}^{\|}$ are equal, as demonstrated in Fig.\ref{fig:KL_plot} for particle loss and dephasing errors, respectively. This is because for these amplitudes $\alpha$ and squeezing degrees $r$ the overlap integral of these states is extremely close to unity. In other words, these states are close to each other on the phase plane. It should also be noted that these SSC states correct particle loss and dephasing errors worse than other types.

Using the KL cost function, we can only compare the encodings and get a qualitative assessment of codewords. For a quantitative assessment of the given QEC protocols, we move on to a more detailed analysis. It is necessary to calculate the channel fidelity $F_{\text{Petz}}$ and construct a recovery procedure $\hat{R}_{\text{Petz}}$. Fig. \ref{fig:fidelity_FPetz} shows the results of the channel infidelity $F_{\text{Petz}}$ for correction of the particle loss (Fig. \ref{fig:fidelity_FPetz}a) and the dephasing errors (Fig. \ref{fig:fidelity_FPetz}b). In the previous section \ref{FPetz_D}, we discussed that in calculating the channel fidelity $F_{\text{Petz}}$ it is necessary to find a recovery operation $\hat{R}_{\text{Petz}}$ given by the projector $\hat{P}_{L}$. It allows us to correct small errors in the quantum channel. Fig. \ref{fig:fidelity_FPetz} illustrates the results of the channel fidelity $F_{\text{Petz}}$. We can see that the recovery operation in the particle loss and dephasing error correction protocol with the second SF state allows achieving channel fidelity of the same order as protocols based on SSC states. It should be noted that $F_{\text{Petz}}$ provides a lower bound on the optimal channel fidelity, i.e., an upper bound on the optimal channel infidelity. Thus, the obtained high value of $F_{\text{Petz}}$ allows us to conclude the high value of the optimal channel fidelity $F_{\text{opt}}$.

Fig. \ref{fig:fidelity_FPetz}a illustrates that the recovery operation based on the second SF state is not optimal in comparison to the considered ones. Indeed, in Fig. \ref{fig:fidelity_FPetz}a, the corresponding channel infidelity curve for the SF state (marked with the green curve) practically coincides with the ones for SSC states with parameters $\alpha_{0.5}^{\perp}$ (marked with the orange dotted curve) and $\alpha_{1.0}^{\perp}$ (marked with the brown curve). The existence of the optimal recovery operation is related to the result of the KL cost function (Fig. \ref{fig:KL_plot}a). One can see that the dependence of the KL cost function on the  $\gamma_{1}$ for the second SF state is lower than the corresponding curves for the SSC states. A similar situation with a "bad"{} recovery operation is observed for the SSC state with $\alpha_{1.0}^{\perp}$ during dephasing error correction. Fig. \ref{fig:fidelity_FPetz}b shows that for this state the dependence of the channel fidelity on the error rate $\gamma_{2}$ is between the curves for the second SF state (marked with the green curve) and the SSC state $\alpha_{0.5}^{\perp}$ (marked with the orange dotted curve). However, the KL cost function curve for the dephasing error for this SSC state, shown in Fig. \ref{fig:KL_plot}b, is less than all other corresponding curves. Also, Fig. \ref{fig:fidelity_FPetz}b demonstrates that the recovery operation for dephasing error correction in the protocol using the second SF state (marked with the green curve) is more optimal than the corresponding recovery operations based on other considered states.

It should be noted that SF states are significantly more advantageous in experimental implementation. Thus, in \cite{PhysRevA.109.052428} a protocol for generating SF states with fidelity equal to 1 was proposed. In turn, the generation of SSC states is possible only approximately. To develop the QEC protocols, we used idealized SSC states. 
Channel fidelity values of the same order were obtained for QEC protocols using the types of states that were considered. This fact is an indisputable confirmation of the competitiveness of SF states in protecting quantum information in a channel with particle loss and dephasing errors.

\section{Conclusion}
In this work, we addressed the problem of quantum error correction in a quantum channel with particle loss and dephasing errors.
We demonstrated an approximate quantum error correction protocol based on the second squeezed Fock states. These states have a certain parity and a special structure in the phase space. For this reason, we considered them as the main resource for quantum error correction. We showed that the second squeezed Fock state is the most appropriate one of the squeezed Fock state set for protecting information in a channel with particle loss and dephasing errors.

We compared the proposed protocol to one based on squeezed Schrödinger cat states. To do this, we used the KL cost function in terms of Kraus operators. This comparison allowed us to obtain preliminary information that the code based on the second squeezed Fock state is better suited for information protection in a channel with particle loss. At the same time, the squeezed Schrödinger cat state with the amplitude $\alpha=1.0$ and the squeezing parameter $ r=1.29$ corrects the dephasing error better than the second squeezed Fock state.

To compare quantum error correction protocols based on the considered states, the Petz map fidelity was also used. We have demonstrated that the order of channel fidelity is the same for both quantum error correction protocols.

In addition, the investigation of the KL cost function and Petz map fidelity for quantum error correction protocols allowed us to identify the general behavior of these measures. Their results are different. Firstly, the KL cost function considers the properties of codewords. Secondly, it does not specify an explicit form for the recovery operation. The result of this measure suggests the existence of an optimal recovery operation. On the other hand, the Petz map fidelity is based on a fixed recovery operation that can correct small errors. In this case, the recovery operation can not be optimal for the given state. The main advantage of the Petz map fidelity is that the recovery operation that is found this way can always be implemented experimentally \cite{PhysRevResearch.6.043034}. However, the issue of transpiling the recovery procedure for the error correction protocol constructed using squeezed Fock states requires a separate study.

In this work, we mainly focused on comparing the error correction code we propose with the code on the squeezed Schrödinger cat states. At the same time, we would like to recall once again the actively developing direction of quantum correction codes on GKP states. Here, significant progress has been achieved in correcting photon loss errors \cite{harris2025logicalchannelheraldedpure,zheng2024performanceachievableratesgottesmankitaevpreskill}. In particular, in \cite{zheng2024performanceachievableratesgottesmankitaevpreskill} it is shown that for a photon loss error rate in a channel of the order of $10^{-1}$, the infidelity reaches a value of the order of $10^{-6}$. However, this code is not applicable to correcting dephasing errors \cite{Joshi_2021,PhysRevA.93.012315}, which forces us to look for worthy alternatives for channels with such losses. It is also important to mention that the GKP codes have made significant progress in terms of scaling to multimode systems \cite{PRXQuantum.5.010331,PRXQuantum.2.030325,PhysRevA.101.012316,PhysRevX.8.021054,matsuura2024}. The limitations and difficulties of similar scaling for the code based on SF states remain beyond the scope of this study.  For the communication channels, it requires an assessment of the rate of the required resource generation. Proof of the ability to perform entangled operations on logical code words is required for quantum error correction in quantum computing.

This research was supported by the Russian Science Foundation (Grant No. 24-22-00004).

%\section*{Disclosures}
%The authors declare no conflicts of interest.
%
%\section*{Data availability} 
%Data underlying the results presented in this paper are not publicly available at this time, but may be obtained from the authors upon reasonable request.

\bibliography{nongaussian}
\appendix
\section{Quantum optimal channel fidelity in terms of the Choi matrix} \label{Fidelity_channel}
Within this Appendix, we would like to give a technique for the derivation of the optimal channel fidelity given in Eq. (\ref{Fopt2}) in terms of the Choi matrix, following the notation of Ref. \cite{noh2021quantumcomputationcommunicationbosonic}.

Let us consider the definition of the general channel presented in Section \ref{general_A}:$\mathcal{Q}= \mathcal{R}\circ\mathcal{N}\circ\mathcal{C}$. Let us consider the definition of the general channel presented in Section \ref{general_A}. Using the definition of the entanglement fidelity, we can write
\BY
F_{e}(\mathcal{Q})\equiv \bra{\Phi^{+}}(\mathcal{Q}\otimes \mathcal{I}_{\mathcal{H}_{e}})(\ket{\Phi^{+}}\bra{\Phi^{+}})\ket{\Phi^{+}},
\label{Fe}
\EY
where $\ket{\Phi^{+}}$ is the maximally entangled state between the system $\mathcal{H}$ and an ancillary system
$\mathcal{H}_{e}$ of the same dimension. This state can be written as
\BY
\ket{\Phi^{+}}=\frac{1}{\sqrt{2}}(\ket{0_{\mathcal{H}}}\ket{0_{\mathcal{H}_{e}}}+\ket{1_{\mathcal{H}}}\ket{1_{\mathcal{H}_{e}}}).
\EY

Note that
\BY
&(\mathcal{Q}\otimes \mathcal{I}_{\mathcal{H}_{e}})(\ket{\Phi^{+}}\bra{\Phi^{+}})=\frac{1}{2}\sum_{\mu,\nu=0}^{1}
\mathcal{Q}(\ket{\mu_{\mathcal{H}}}\bra{\nu_{\mathcal{H}}})\otimes\ket{\mu_{\mathcal{H}_{e}}}\bra{\nu_{\mathcal{H}_{e}}}
=\\\nonumber&=\frac{1}{2}\sum_{\mu,\nu=0}^{1}\sum_{\rho,\sigma=0}^{1}[\textbf{X}_{\mathcal{Q}}]_{[\mu \rho],[\nu \sigma]}(\ket{\rho_{\mathcal{H}}}\bra{\sigma_{\mathcal{H}}})\otimes\ket{\mu_{\mathcal{H}_{e}}}\bra{\nu_{\mathcal{H}_{e}}},
\label{adeq}
\EY
where $[\textbf{X}_{\mathcal{Q}}]_{[\mu \rho],[\nu \sigma ]}\in \mathcal{B}(\mathcal{H}\otimes\mathcal{H})$ is the Choi matrix \cite{Schlegel_2022,Kosut2009} of the channel $\mathcal{Q}$. The Choi matrix $[\textbf{X}_{\mathcal{Q}}]_{[\mu \rho],[\nu \sigma]}$ fully characterizes the channel $\mathcal{Q}$, similar to the set of Kraus operators $\{\hat{Q}_{j}\}$. The complete positivity condition of the map $\mathcal{Q}$ corresponds to the fact that $[\textbf{X}_{\mathcal{Q}}]_{[\mu \rho],[\nu \sigma]}\succeq 0$. At the same time, the trace-preserving condition for the map $\mathcal{Q}$ in terms of the Choi matrix has the form
\BY
\Tr_{\mathcal{H}_{e}}[\textbf{X}_{\mathcal{Q}}]\equiv \sum_{\rho=0}^{\dim(\mathcal{H}_{e})}[\textbf{X}_{\mathcal{Q}}]_{[\mu \rho],[\nu \sigma]}\ket{\mu_{\mathcal{H}}}\bra{\nu_{\mathcal{H}}}=\mathcal{I}_{\mathcal{H}}.
\label{cond2}
\EY
Here, $\mathcal{I}_{\mathcal{H}_{e}}$ is the identity channel acting on the $\mathcal{H}_{e}$ ancillary system. Note that the positive semidefinite of the Choi matrix $[\textbf{X}_{\mathcal{Q}}]_{[\mu \rho],[\nu \sigma]}$ and the requirement (\ref{cond2}) lead to the parameters of the Choi matrix $[\textbf{X}_{\mathcal{Q}}]_{[\mu \rho],[\nu \sigma]}$  form a convex set.

Using the expression (\ref{adeq}), we can simplify the expression for the fidelity  Eq.(\ref{Fe}) as 
\BY
F_{e}(\mathcal{Q})=\frac{1}{2}\bra{\Phi^{+}}\big(\sum_{\mu,\nu=0}^{1}\sum_{\rho,\sigma=0}^{1}[\textbf{X}_{\mathcal{Q}}]_{[\mu \rho],[\nu \sigma]}(\ket{\rho_{\mathcal{H}}}\bra{\sigma_{\mathcal{H}}})\otimes\ket{\mu_{\mathcal{H}_{e}}}\bra{\nu_{\mathcal{H}_{e}}}\big)\ket{\Phi^{+}}=\frac{1}{4}\sum_{\mu,\nu=0}^{1}[\textbf{X}_{\mathcal{Q}}]_{[\mu \mu],[\nu \nu]},
\label{Fe2}
\EY
where the matrix elements of the Choi matrix are defined as
\BY
[\textbf{X}_{\mathcal{Q}}]_{[\mu \rho],[\nu \sigma]}=\bra{\rho_{\mathcal{H}}}\mathcal{Q}(\ket{\mu_{\mathcal{H}}}\bra{\nu_{\mathcal{H}}})\ket{\sigma_{\mathcal{H}}}.
\EY

Similarly, we can define Choi matrices for the encoding $\mathcal{C}$, the noise $\mathcal{N}$, and the recovery $\mathcal{R}$ maps:
\BY
&[\textbf{X}_{\mathcal{C}}]_{[\mu \rho],[\nu \sigma]}=\bra{\rho_{\mathcal{H}_{d}}}\mathcal{C}(\ket{\mu_{\mathcal{H}}}\bra{\nu_{\mathcal{H}}})\ket{\sigma_{\mathcal{H}_{d}}},
\\&[\textbf{X}_{\mathcal{N}}]_{[\mu \rho],[\nu \sigma]}=\bra{\rho_{\mathcal{H}_{d}}}\mathcal{N}(\ket{\mu_{\mathcal{H}_{d}}}\bra{\nu_{\mathcal{H}_{d}}})\ket{\sigma_{\mathcal{H}_{d}}},
\\&[\textbf{X}_{\mathcal{R}}]_{[\mu \rho],[\nu \sigma]}=\bra{\rho_{\mathcal{H}_{d}}}\mathcal{R}(\ket{\mu_{\mathcal{H}_{n}}}\bra{\nu_{\mathcal{H}_{n}}})\ket{\sigma_{\mathcal{H}_{d}}},
\EY
where $\mu,\nu,\rho,\sigma\in\{0,1\}$.

Using $\mathcal{Q}= \mathcal{R}\circ\mathcal{N}\circ\mathcal{C}$, we can transform the Eq.(\ref{Fe2}) as
\BY
F_{e}(\mathcal{Q})=\frac{1}{4}\sum_{i,j=0}^{1}[\textbf{X}_{\mathcal{R}\circ\mathcal{N}\circ\mathcal{C}}]_{[i i],[j j]}=\frac{1}{4}\sum_{i,j=0}^{1}[\textbf{X}_{\mathcal{R}\circ\mathcal{N}\circ\mathcal{C}}]_{i i,j j}.
\label{Fe3}
\EY

Next, to further transform the expression of the optimal channel fidelity, we consider the superoperator $\hat{T}_{\mathcal{A}}$ of the quantum map $\mathcal{A}: \mathcal{B}(\mathcal{H}_{1})\rightarrow \mathcal{B}(\mathcal{H}_{2})$. The matrix elements  are defined as $[\hat{T}_{\mathcal{A}}]_{\mu \rho,\nu \sigma}\equiv [\textbf{X}_{\mathcal{A}}]_{[\mu \rho ],[\nu \sigma]}$. We should note an important property of the superoperator:
\BY
\hat{T}_{\mathcal{B}\circ\mathcal{A}}=\hat{T}_{\mathcal{B}}\hat{T}_{\mathcal{A}}.
\label{suoop}
\EY
Then we can transform the  Eq.(\ref{Fe3}) as follows:
\BY
F_{e}(\mathcal{Q})=\frac{1}{4}\sum_{i,j=0}^{1}[\textbf{X}_{\mathcal{R}\circ\mathcal{N}\circ\mathcal{C}}]_{[i i],[j j]}=\frac{1}{4}\sum_{i,j=0}^{1}[\hat{T}_{\mathcal{R}\circ\mathcal{N}\circ\mathcal{C}}]_{i j,i j}=\frac{1}{4}\Tr[\hat{T}_{\mathcal{R}\circ\mathcal{N}\circ\mathcal{C}}],
\label{Fe4}
\EY
where $\Tr$ in this case denotes the trace of the superoperators.

Given that the property of the superoperator Eq.(\ref{suoop}) to simplify the Eq. (\ref{Fe4}):
\BY
[\hat{T}_{\mathcal{R}\circ\mathcal{N}\circ\mathcal{C}}]_{j j^{'},i i^{'}}=\sum_{n,n^{'},m,m^{'}=0}^{1}[\textbf{X}_{\mathcal{R}}]_{[m j],[m^{'} j^{'}]}[\textbf{X}_{\mathcal{N}}]_{[n m],[n^{'} m^{'}]}[\textbf{X}_{\mathcal{C}}]_{[i n],[i^{'} n^{'}]},
\EY
for $j,j^{'},i,i^{'}\in\{0,1\}$,  where $\textbf{X}_{\mathcal{R}}\in\mathcal{B}(\mathcal{H}_{d}\otimes \mathcal{H}_{n})$, $\textbf{X}_{\mathcal{N}}\in\mathcal{B}(\mathcal{H}_{n}\otimes \mathcal{H}_{d})$, $\textbf{X}_{\mathcal{C}}\in\mathcal{B}(\mathcal{H}_{d}\otimes \mathcal{H})$ are Choi matrices of the recovery $\mathcal{R}$, noise $\mathcal{N}$ and  encoding $\mathcal{C}$ maps, respectively. Next, we can write the process matrix in terms of Choi matrices \cite{Schlegel_2022,Kosut2009} as 
\BY
[\textbf{W}_{\mathcal{R}}]_{[m^{'} i^{'}],[m i]}=\sum_{n,n^{'}=0}^{1}[\textbf{X}_{\mathcal{N}}]_{[n m],[n^{'} m^{'}]}[\textbf{X}_{\mathcal{C}}]_{[i n],[i^{'} n^{'}]},
\label{w}
\EY
where $m,m^{'}\in\{0,1\}$. The process matrix $\textbf{W}_{\mathcal{R}}$ can also be written in terms of Kraus operators \cite{Schlegel_2022}, as can be seen from Eq.(\ref{Fopt_m2}) in section \ref{general_B}.
Thus, the channel fidelity in terms of Choi matrices can be represented as
\BY
F_{e}(\mathcal{Q})=F_{e}(\mathcal{R}\circ\mathcal{N}\circ\mathcal{C})=\frac{1}{4}\Tr[\textbf{X}_{\mathcal{R}}\textbf{W}_{\mathcal{R}}].
\label{Fe5}
\EY

\section{A selection of codewords from a set of squeezed Fock states}  \label{SF_codewords}
In this Appendix, we would like to discuss the choice of the optimal number $n$ of SF states that encode the logical states $|0_{L}\rangle$ and $|1_{L}\rangle$.  To do this, the SF states should be orthogonal. At the same time, to simplify the experimental implementation of the error correction code, these states should have a minimum number of $n$.

Consider the state $|r,0\rangle$ with the number $n=0$. The scalar product of the states $|r,0\rangle$ and $|-r,0\rangle$ has the form
\begin{align}
  \langle -r,0  |r,0\rangle= \int_{-\infty}^{\infty} \Psi^{*}_{\mathrm{\mathrm{SF}}}(x,-r,0)\Psi_{\mathrm{\mathrm{SF}}}(x,r,0)\;dx=\frac{\sqrt{2}e^{r}}{\sqrt{1+e^{4r}}},
  \label{state_n=0}
\end{align}
where the wave function of the SF state \cite{PhysRevA.109.052428} has the following form
\begin{align}
	\label{SFS}
	\Psi_{\mathrm{\mathrm{SF}}}(x, r, n)=\frac{e^{-\frac{1}{2} e^{2 r} x^2} H_n\left(e^r x\right)}{\sqrt{2^n n! \sqrt{\pi} e^{-r}}}.
\end{align}
Here, $r$ is the squeezing parameter. Eq. (\ref{state_n=0}) shows that there is no value of the squeezing parameter $r$ that achieves orthogonality of the states $|r,0\rangle$ and $|-r,0\rangle$. For this reason, these states are not optimal as codewords and are not right for us.

Consider the state $|r,1\rangle$ with the number $n=1$. The corresponding scalar product can be written as
\begin{align}
  \langle -r,1  |r,1\rangle= \int_{-\infty}^{\infty} \Psi^{*}_{\mathrm{\mathrm{SF}}}(x,-r,1)\Psi_{\mathrm{\mathrm{SF}}}(x,r,1)\;dx=\frac{1}{\cosh(2r)^{3/2}}.
\end{align}
Here, we can also see that there is no value for the squeezing parameter $r$ that achieves orthogonality of the corresponding states. Consequently, we shouldn't use such states as codewords.

Let us consider the state $|r,2\rangle$ with the number $n=2$. The corresponding scalar product has a form
\begin{align}
  \langle -r,2  |r,2\rangle= \int_{-\infty}^{\infty} \Psi^{*}_{\mathrm{\mathrm{SF}}}(x,-r,2)\Psi_{\mathrm{\mathrm{SF}}}(x,r,2)\;dx=-\frac{\sqrt{2}e^{5r}(\cosh(4r)-5)}{(1+e^{4r})^{5/2}}.
  \label{SF_n=2}
\end{align}

As we can see that for $r=\pm \arccosh(5)/4\approx \pm 0.57$, the orthogonality of the states $|r,2\rangle$ and $|r,2\rangle$ is achieved. Thus, the second SF state with the squeezing parameter $r=0.57$ should be used as codewords. For SF states with $n>2$, there is the squeezing parameter $r$ such that SF states are orthogonal. Nevertheless, the probability of generating such states will decrease as the number $n$ increases \cite{PhysRevA.109.052428}. As a result, the probability of generating the second SF state is $14.81\%$, the third one is $10.55\%$, and the fourth one is already $8.19\%$. For this reason, we will consider only the second SF states as the codewords. It is worth noting that the scheme for generating SF states \cite{PhysRevA.109.052428} is a scheme with notification. Therefore, the state generation probabilities considered above are the probabilities of the notification being triggered. If the notification being triggered, then we obviously have the required resource and can use it in the error correction scheme. We do not use states in the protocol that are not suitable for correction.

\end{document}